\documentclass[twocolumn,showpacs,preprintnumbers]{revtex4}
\usepackage{amssymb}
\usepackage{amsmath}
\usepackage{graphicx}
\usepackage{epsfig}

\begin{document}
\title{\bf  Possible two-gap superconductivity in NdFeAsO$_{0.9}$F$_{0.1}$ probed by point-contact Andreev-reflection spectroscopy}

\author{P. Samuely,$^{1}$ P. Szab\'o,$^{1}$ Z. Pribulov\'a,$^{1}$ 
M. E. Tillman$^{2}$, S. Bud'ko,$^{2}$ and P. C. Canfield$^{2}$ }

\affiliation{$^1$Centre of  Low Temperature Physics, IEP Slovak
Academy of Sciences  \&  P.J.\v Saf\'arik University, Watsonova
47, SK-04001 Ko\v sice, Slovakia\\
$^2$Ames Laboratory and Iowa State University, Ames, IA 50011, USA.
}

\begin{abstract}

Systematic studies of the NdFeAsOF superconducting energy gap via the point-contact Andreev-reflection (PCAR) spectroscopy are presented. The PCAR conductance spectra show  at low temperatures a pair of gap-like peaks at about $\pm (4 \div 7)$ mV indicating the superconducting energy gap and in most cases also a pair of humps at around $\pm $10 mV.  Fits to the $s$-wave two-gap model of the PCAR conductance allowed to determine two superconducting energy gaps in the system. The energy-gap features  however disappear already at $T^* = 15$ to 20 K, much below the particular $T_c$ of the junction under study. At $T^*$ a zero-bias conductance (ZBC) peak emerges, which at higher temperatures usually overwhelms the spectrum with intensity significantly higher than the conductance signal at lower temperatures. Possible causes of this unexpected temperature effect are discussed. In some cases the conductance spectra show just a reduced conductance around the zero-bias voltage, the effect persisting well above the bulk transition temperature. This indicates a presence of the pseudogap in the system.

\end{abstract}

\maketitle




\section{INTRODUCTION}
Although the field of superconductivity near its centenary anniversary has been still extremely 
vigorous the discovery of high temperature superconducting iron-based pnictides  is a significant breakthrough \cite{kamihara}. REFeAsO(F) systems with various rear earths (RE) elements  bring a new class of layered high-$T_c$ materials with numerous similarities with high-$T_c$ cuprates, from antiferromagnetism in parent compounds (albeit metallic), through electron and hole doping as a route leading to superconductivity, to  a possible unconventional pairing mechanism. NdFeAsO$_{0.9}$F$_{0.1}$ with $T_c$ above 51 K  \cite{ren} together with the Sm and Pr compounds reveal the highest transition temperature at ambient pressure among all pnictides. One of the most fundamental issues for unveiling the superconducting mechanism of these multiband systems concerns the symmetry of the superconducting order parameter(s). There are numerous theoretical predictions  on this topic but also a body of  experimental studies is emerging. Band structure calculations have shown disconnected sheets of the Fermi surface with possibly different superconducting energy gaps. A minimal model has to include two bands: the hole band around the $\Gamma$ point and the electron one around the $M$ point \cite{scalapino}. In contrast to the multiband but conventional $s$-wave scenario in MgB$_2$ \cite{szabo}, here the extended $s$-wave pairing with a sign reversal of the order parameter between different Fermi surface sheets has been proposed by Mazin {\it et al.} \cite{mazin}. The iron pnictides would be the first example of a multigap superconductor with a discontinuous sign change of the order-parameter phase between bands.    Li and Wang \cite{li} have proposed that the pairing occurs in
the {\it d}-wave channel where, by lowering temperature,
the system enters first the $d_{xy}$ superconducting phase and then enters the
time-reversal-symmetry-broken $d_{xy} + id_{x^2-y^2}$ superconducting phase.      

Early experimental results are starting to provide some insight as well. The $H_{c1}$ magnetization measurements on F-doped LaFeAsO \cite{ren2} suggest a nodal gap function showing  a clear linear temperature behavior at low temperatures. Matano {\it et al.} \cite{matano} in their NMR studies on PrFeAsOF found two superconducting energy gaps but in contrast to the case of MgB$_2$ here with nodes. The penetration depth studies have provided the nodeless superconducting energy gap in NdFeAsO$_{0.9}$F$_{0.1}$ with remarkably small coupling 2$\Delta/k_BT_c \approx $2 \cite{martin}. 

The point-contact Andreev reflection (PCAR) spectroscopy has been very powerful technique in investigating the superconducting order parameter even in case of  multiple gaps like in MgB$_2$ \cite{szabo}. The PCAR spectroscopy data known on pnictides so far have brought conflicting results. Shan {\it et al.} \cite{shan} found a very pronounced zero-bias conductance peak ascribed to the Andreev bound states ($d$- or $p$-wave pairing) accompanied by weak signatures of the superconducting energy gap indicating the weak coupling with 2$\Delta/k_BT_c$ equal to about 3.5. Chen {\it et al.} \cite{tesanovic} have presented a surprisingly conventional superconducting energy gap with a medium coupling equal to 2$\Delta/k_BT_c \approx 3.7$. The recent data of  Yates {\it et al.} \cite{cohen} obtained on the 45 K NdFeAsO$_{0.85}$ show also an indication of the superconducting energy gap with 2$\Delta/k_BT_c = 3.6$. 

In the following we present the systematic PCAR studies on the NdFeAsO$_{0.9}$F$_{0.1}$ polycrystals indicating a two-gap $s$-wave superconductivity in the system. In the conductance spectra the smaller gap is pronounced in the form of two peaks positioned symmetrically at $\pm  (4\div 7)$ mV while the large one is revealed at the humps at about $\pm  10$ mV. In some cases the point-contact spectra show no signature of enhanced conductance due to the Andreev reflection of quasiparticles on the superconducting energy gap. In this case just a reduced conductance near the zero-bias voltage is observed persisting even well above the bulk transition temperature. This indicates a presence of the pseudogap in the system. Surprisingly, the gap features   disappear from the spectra already at $T^* =15\div 20$ K, much below the particular $T_c$ of the junction under study. At $T^*$ the spectrum is reduced to a zero-bias peak with intensity sometimes significantly higher than the conductance signal at lower temperatures. Possible reasons for this temperature behavior are discussed.

\section{EXPERIMENT}

Samples with a nominal
composition NdFeAsO$_{0.9}$F$_{0.1}$ have been prepared by the high pressure synthesis in a cubic,
multianvil apparatus, with an edge length of 19
mm from Rockland Research Corporation. Stoichiometric
amounts of NdFe$_3$As$_3$, Nd$_2$O$_3$, NdF$_3$ and Nd were
pressed into a pellet with mass of approximately 0.5 g
and placed inside of a BN crucible with an inner diameter
of 5.5 mm. The synthesis was carried out at about
3.3 GPa. The temperature was increased over a period
of one hour to 1350 - 1400 $^o$C and held for 8 hours before
being quenched to room temperature. The pressure
was then released and the sample removed mechanically.
More details of the synthesis and characterization will be
found elsewhere \cite{tillman}. The value of 10\% F substitution
is nominal, based on the initial stoichiometry of the pellet.
The synthesis yields polycrystalline NdFeAsO$_{0.9}$F$_{0.1}$
samples that contain what appears to be plate-like single
crystals as large as 300 $\mu$m \cite{prozorov}. Whereas extraction of
these crystallites is difficult, we could measure properties
of individual crystals by using local point-contact probe with a metallic tip.
   
The point-contact Andreev-reflection (PCAR)
measurements  have been  realized via  the standard  lock-in
technique in a special point-contact approaching system with
lateral and vertical movements of the PC tip by a differential
screw  mechanism.  The  microconstrictions  were prepared in
situ  by  pressing  different   metallic  tips  (copper  and
platinum  formed either  mechanically or  by electrochemical
etching) on different parts  of the freshly polished surface
of  the  superconductor. Typical dimension of the microconstriction formed by such method is usually on the order of tens of nanometers. 
$T_c$'s have been determined
from the resistive transitions
and   also  from   the  temperature   dependences  of   the
point-contact spectra. While onset of the transition to the zero-resistance state was at 51 K the  transition temperatures found by
the point-contact technique varied between 45 and 51 K.

The point-contact    spectrum   measured    on   the    ballistic
microconstriction   between   a   normal   metal   (N)   and
a superconductor (S) consist of  the Andreev reflection (AR)
contribution and the  tunneling contribution \cite{btk}. The
charge    transfer    through    a    barrierless   metallic
point contact  is  realized  via  the  Andreev reflection of
carriers. Consequently  at $T = 0 $  the PC current as  well as
the PC  conductance inside the gap  voltage ($V< \Delta /e$)
is  twice  higher  than  the  respective  values  at  higher
energies  ($V>>\Delta /e$).  The presence  of the  tunneling
barrier  reduces the  conductance at  the zero  bias and two
symmetrically  located peaks  rise  at  the gap  energy. The
evolution  of  the  point-contact  spectra  between the pure
Andreev reflection  and the Giaever-like  tunneling has
been       theoretically        described       by       the
Blonder-Tinkham-Klapwijk   (BTK)   theory   \cite{btk} for the case of $s$-wave superconductors.  The
point-contact  conductance data  can be  compared with  this
theory using  as input parameters  the energy gap  $\Delta$,
the parameter $z$ (measure for the strength of the interface
barrier)   and  a   parameter  $\Gamma$   for  the  spectral
broadening \cite{plecenik}. In  any case the  voltage dependence of  the N/S
point-contact   conductance   gives   direct   spectroscopic
information   on   the   superconducting   order   parameter
$\Delta$.  For   a  multiband/multigap superconductor the
point-contact conductance $G = dI/dV$ can be expressed as a weighted
sum of  partial BTK  conductances. As shown in our previous work \cite{szabo} in case of MgB$_2$ the total PCAR conductance could be simply summarized  from two contributions of band 1 and 2 (the 3D  $\pi$-band with a small
gap $\Delta_{1}$ and the quasi
two-dimensional  $\sigma$-band  with  a  large  gap $\Delta
_{2}$, respectively)
\begin{equation}
G  =  \alpha  G_{1}  +  (1-\alpha )G_{2}.
\end{equation}
where $\alpha$ is the weight factor for  the  $ \pi$ - band 
contribution \cite{szabo}. 
Tanaka {\it et al.} \cite{tanaka} have extended the BTK  formulation also for unconventional pairing 
($d$-wave or $p$-wave).

\section{RESULTS AND DISCUSSION}

Before evaluating the point-contact spectra precautions have to be made that the heating effects are avoided and the junction is in a ballistic or spectroscopic regime. One way is to evaluate the contact size and check if it is smaller compared to the quasiparticle mean free path in the normal as well as in the superconducting electrodes and superconducting coherence length.  Even if the bulk resistivities of the tip and the sample are known (which is not the case for our NdFeAsO$_{0.9}$F$_{0.1}$) the local resistivity at the point-contact area can be significantly enhanced due to a pressure of the tip on the sample. Then, the evaluation of the contact size and the quasiparticle mean free path becomes problematic. Moreover very often multiple parallel contacts are formed.  That is why inspection of typical heating effects is usually used to determine whether the spectroscopic regime is established in the junction. In the following we present the data where the heating effects, as sharp conductance dips and irreversibilities in the conductance curves usually observed at low voltages, are absent. Moreover, we have focused on the junctions with important tunneling component indicated by the interface barrier-strength parameter $z > 0$. 

Figure 1 shows a variety of the PCAR spectra obtained on the junctions made by Pt tip on the NdFeAsO$_{0.9}$F$_{0.1}$ polycrystalline samples at 4.3 K (solid lines).  The spectra labeled as A, B, C and D have been normalized to their particular normal-state conductances measured above $T_c$. The spectrum E was just divided by its own value at 60 mV. For the sake of clarity the lower curves are vertically shifted. As can be seen the spectra  show a pair of the gap-like peaks symmetrically placed at $\sim  \pm(5)$ mV (the curves A, B and C) or at $\sim  \pm(7)$ mV in the case of the spectrum D. In the spectra A,B and C also a reproducible shoulder near $\sim$ 10 mV and some smaller structures with much less reproducibility are revealed.  

Location of the gap-like peaks/humps is indicative for a size of the related gap but precise determination of the gap is not possible in this way. For example in the case of important spectral broadening, typically found in samples of new and in addition nonstoichiometric materials, the gap-like peaks can be pushed to significantly higher voltages than would correspond to a real size of the gap. Another uncertainty can be caused  by a way of normalization of the spectra, etc. One way to partly overcome this troubles is to compare the measured data to the appropriate model. In the following the spectra showing apparent enhanced conductance and the gap-like peaks (both features indicating the Andreev reflection of quasiparticles on the superconducting energy gap) have been fitted to the BTK single-gap formula. The resulting fits except for the curve D were poor and unable to reproduce the measured curves. The failure did not concern only the humps at about 10 mV but also the overall width of the enhanced conductance around the zero-bias was impossible to achieve. Then, in the case of the spectra A, B, and C we proceeded to the two-gap BTK model mentioned above. As can be seen from the curves indicated by the open circles in Fig. 1, the two gap $s$-wave BTK formula represents a good fit to the data. We remark that even if the two-gap model describes  our spectra measured at $T$ = 4.3 K quite well, small deviations occur at higher voltages above $\Delta_2$ value.  
The following fitting parameters have been obtained: 
A - $\Delta_1$ = 5 meV, $\Delta_2$ = 12 meV. B - $\Delta_1$ = 5.2 meV, $\Delta_2$ = 10 meV. C - $\Delta_1$ = 5 meV, $\Delta_2$ = 12 meV.
The smearing parameters $\Gamma_{1,2}$ vary but they have always been in the range $\Gamma_{i}<0.4 \Delta_{i}$ of the respective gap value. The weight factor $\alpha $ for contribution of the band 1 with a smaller gap $\Delta_1$ has been scattered between 0.4 and 0.6. In the case of the spectrum D the fit by the single-gap formula was possible yielding  $\Delta_1 =4 $ meV with a significant large smearing $\Gamma = 3.35 $ meV. The barrier strength parameter $z$  has been found between 0.3 and 0.5 in the case of the curves A, B, C and D as well as in the case of the junctions with the similar gap-like spectra. 

The particular transition temperature of the junctions A,B, C and D was very close to 45 K. It is documented by the temperature dependence of the spectrum B displayed in Fig. 3a, where the enhanced conductance around zero bias is completely gone at this temperature. Taking a value of $T_c = 45$ K the following superconducting coupling strengths are obtained from the junction A, B and C: $2 \Delta_1 /k_B T_c = 2.6 \pm 0.1$ and 
$2 \Delta_2 /k_B T_c= 5.7 \pm 0.5 $. Consequently, a weak coupling below the canonical BCS single-band value is found in the band 1 with the small gap and a strong coupling is characteristic for the second band. In the case of the junction D the coupling strength is even smaller with $2 \Delta_1 /k_B T_c = 2.1$ indicating that probably the information is not complete.  

Inspecting the data obtained on numerous junctions revealing the gap-like features has shown that the values of the small gap $\Delta_1$ vary between 4 and 6 meV and for the large one $\Delta_2$ between 9 and 13 meV.

\begin{figure}[t]
\includegraphics[width=7 cm]{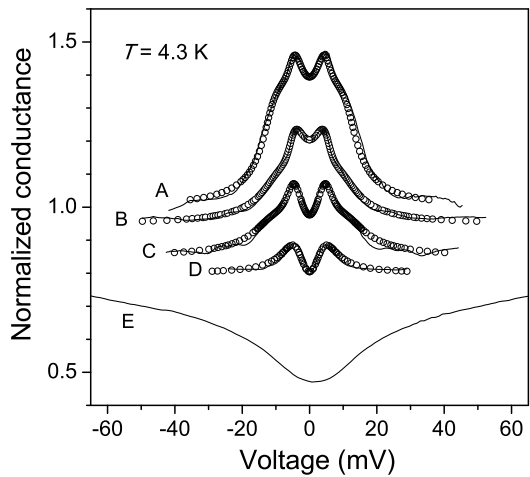}
\caption{Several characteristic PCAR spectra of the point-contact junction between Pt tip and NdFeAsO$_{0.9}$F$_{0.1}$ measured at 4.3 K - solid lines.  The lower spectra are vertically shifted for the clarity. Open symbols represent fitting curves to the $s$-wave two-band BTK model.}
\label{fig:fig1}
\end{figure}

\begin{figure}[t]
\includegraphics[width=7 cm]{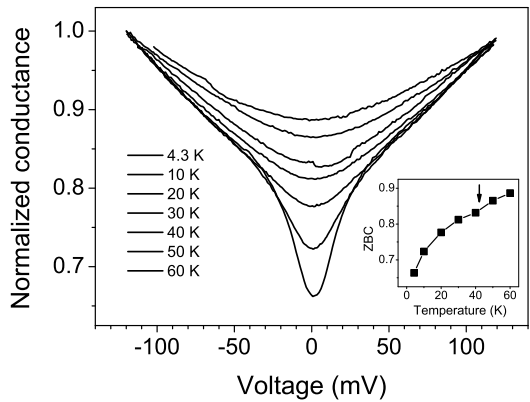}
\caption{Spectra showing a reduced conductance even above $T_c$. The spectra have been normalized to their values at -100 mV. Inset - The zero-bias conductance as a function of temperature, the arrow depicts a position of the critical temperature in the system.}
\label{fig:fig2}
\end{figure}

\begin{figure}[t]
\includegraphics[width=7 cm]{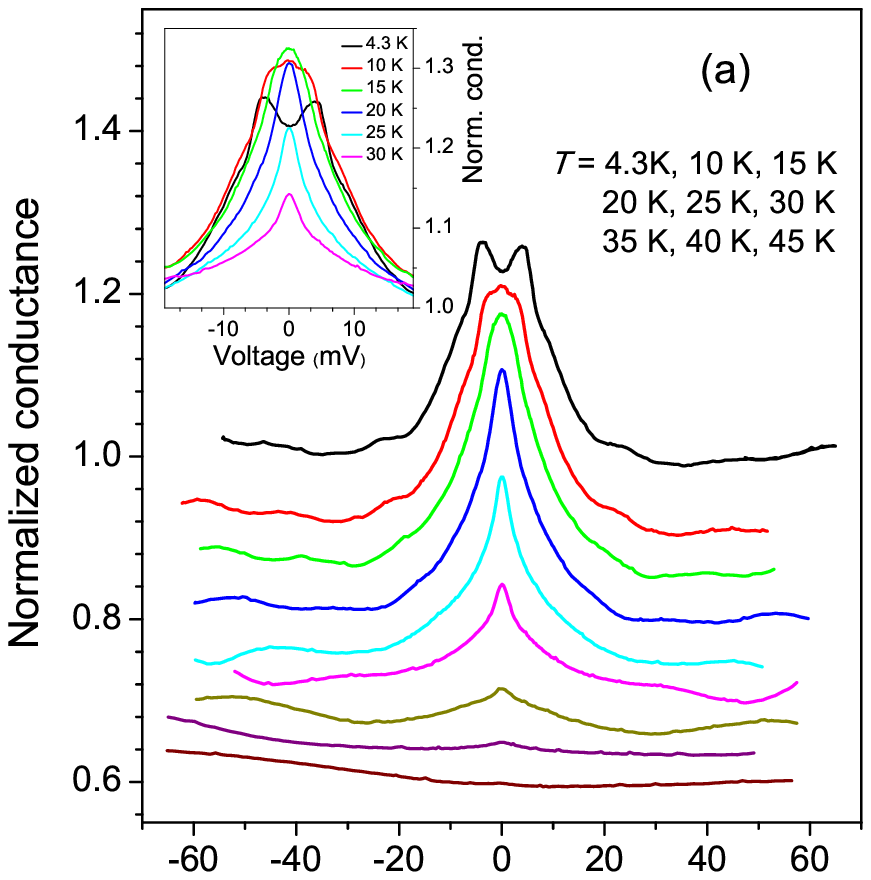}
\label{fig:fig3}
\end{figure}
\begin{figure}
\includegraphics[width=6.5 cm]{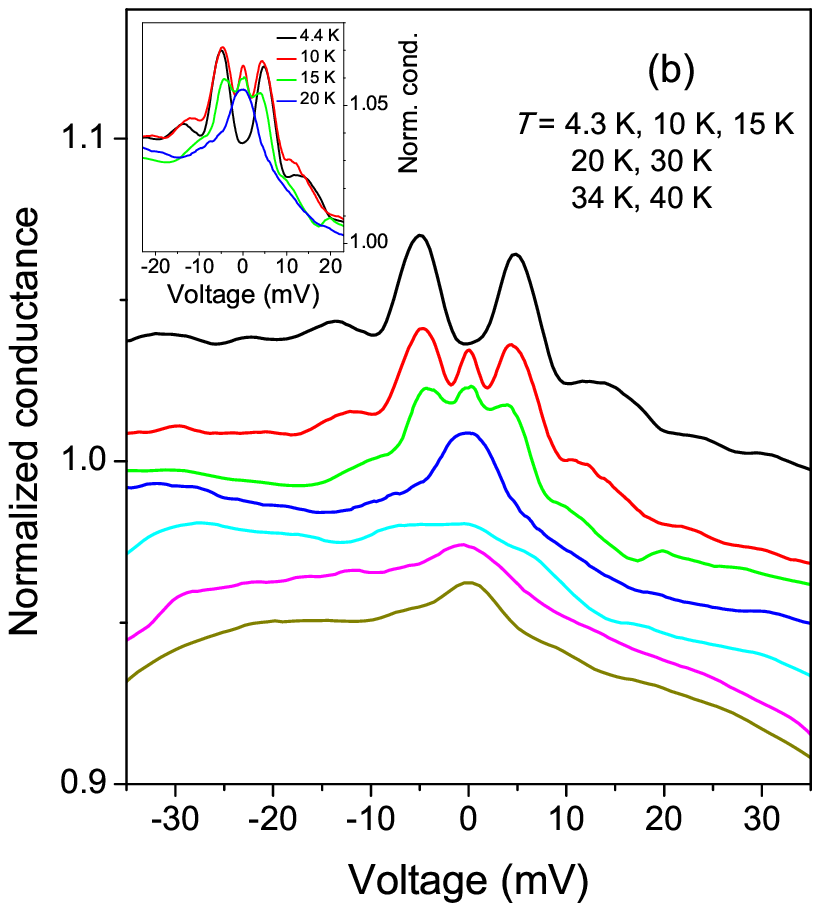}
\caption{PCAR conductance spectra of NdFeAsO$_{0.9}$F$_{0.1}$ showing the superconducting-gap features measured at different temperatures. The lower spectra are vertically shifted. The insets show details of an unusual evolution of the spectra at lower temperatures (see the text).  The spectrum displayed in 3a is the same as the curve B in Fig. 1.}
\label{fig:fig3}
\end{figure}

Quite a few of the measured spectra revealed a behavior indicated at the spectrum E in Fig. 1. These spectra have  shown neither enhanced point-contact conductance nor coherence peaks, the both effects indicating the Andreev reflection of quasiparticles on the superconducting energy gap but they have rather displayed a reduced conductance around the zero-bias voltage. 

In Fig. 2 the effect of temperature on another junction with tunneling-like characteristics is shown. One can notice that the effect of the reduced conductance persists well above the bulk transition temperature $T_c \approx $ 45 K. It is better observable from the {\it ZBC versus temperature} dependence shown in the inset  of Fig. 2. 
Then, the conclusion can be made that this tunneling-like feature can not be connected with the superconductivity in the system but reminds a reduced density of states of the normal quasiparticles or pseudogap observed in the underdoped high-$T_c$ cuprates. Since the measurements have been done on polycrystals one can only speculate if the tunneling-like conductance characteristics are related to a specific direction of the point-contact current with respect to the crystallographic orientation of the sample. In this line it is worth noticing that very recently we have observed this kind of characteristics on the single crystals of (Ba,K)Fe$_2$As$_2$ when the point-contact current was oriented in the $c$ crystallographic direction, while the spectra with two superconducting gaps were observed in the $ab$ planes \cite{szabo2}. 
  
In Fig. 3a and 3b the temperature dependence of the spectra showing the superconducting gap-like features is shown. The presented curves are in fact the raw data just normalized to their values at voltage $V$ = 40 mV. The lower curves are shifted for clarity.
Similarly to all others of the numerous  spectra we have measured, the spectral backgrounds reveal a small asymmetry, being higher at negative bias voltage, i.e. when the electrons are injected from the tip into the superconductor. This asymmetry is revealed also in the spectra with a tunneling-like character as those displayed in Fig. 2. This feature revealed also above $T_c$ is in agreement with the measurements of Chen et al. \cite{tesanovic}. One possible explanation can be that the asymmetry  is due to energy dependent DOS in the normal state.

The both sets of spectra in Fig. 3  reveal at 4.3 K two symmetrically placed  gap-like peaks at $\sim  5$ mV and humps at $\sim  10\div 13$ mV and higher voltage. The effect of temperature is very unusual. Namely, while the position of the gap-like peaks is rapidly shrinking with increasing temperature, surprisingly the intensity of the conductance spectrum near the zero-bias voltage increases. The latter effect is better presented in the insets, where the curves are not vertically shifted. Above 10 K the only visible feature of the spectrum 3a is the narrow ZBC peak which at even higher temperatures looses its intensity and vanishes at the local $T_c$, here at about 45 K. In the second set of the spectra shown in Fig. 3b the ZBC peak appears already at $T$ = 10 K and coexists with the gap-like peaks up to $T$ = 20 K. If one looks to the evolution of the spectrum from higher to lower temperatures, it seems that below $T_c$ the peak in conductance at zero bias is evolving and at certain transition temperature $T^*$ a gap is opening on this central conductance maximum.

We remark that the temperature $T^*$ where the gap-like peaks are replaced by the zero-bias maximum is not always the same but depends on the strength of the interface barrier $z$.  The higher $z$, the higher temperature $T^*$. This finding is consistent with a behavior of the point-contact junctions on SmFeAsO$_{0.85}$F$_{0.15}$ as presented in Fig. 2 of Ref.\cite{tesanovic}. Co-presence of the gap-like peaks and the zero-bias peak has also been found in Ref.\cite{cohen}. There, in Fig. 1 one can  moreover notice that, similarly to our observations, the hight of the ZBC peak at 32 K is larger than the zero-bias conductance at 5.2 K.

What could be an explanation of the observed ZBC peak in the PCAR spectra? Indeed there is an analogy with cuprates where it is  connected with the nodal superconducting energy gap or the order parameter phase sign changing along the (110) surface.  For the iron pnictides Mazin and coworkers \cite{mazin} suggested two $s$-wave gaps on disconnected Fermi surface sheets but with opposite sign. Recently, Choi and Bang \cite{bang} have adopted this model of "$s\pi $" pairing ($s$-wave order parameter with sign reversal) in analogy with a behavior of the superconducting/ferromagnet (S/F) bilayers where also the order parameter sign change can happen.  Their calculations show that in the case of the two-band system the superconductivity is possible even for a negative pairing interaction  by generating the sign reversal between the gaps $\Delta_1$ and $\Delta_2$.  One of the consequencies of the  $s\pi $ state is that one gap is smaller and another one larger than the BCS weak coupling value. 
The calculations have also shown that $s\pi $ state can produce the zero-bias peak in the local density of states at the superconducting/normal interface. This can be detected in the tunneling or PCAR spectroscopy measured on the normal metal/superconductor junctions. Specific conditions enabling observations the ZBC peak must be studied in details to explain why this effect would be appear only at higher temperatures. But it is noteworthy that the ZBC peak effect is observed in the  all PCAR tunneling data available so far \cite{shan,tesanovic,cohen}.

On the other hand the presence of the ZBC peak in the point-contact spectrum could be just a fingerprint of the local destruction of superconductivity, i.e the point contact would rather be in a thermal than in spectroscopic regime. Obviously, even the point contact revealing spectroscopic features in the form of gap-like peaks at low temperatures could be driven out of the spectroscopic regime by higher temperatures and/or higher voltages. This is due to the fact that a mean free path of the quasiparticles is shortened at higher temperatures/voltages. This scenario seems to contradict the behavior seen in Fig. 3b, where the ZBC peak is accompanied by a presence of the gap-like peaks at 10 and 15 K. The higher voltage can not redirect the junction back into the spectroscopic regime. It means that once the junction is in the thermal regime at voltages near the zero bias it can not show up such a pronounced gap-like peaks at higher voltages. But we can not completely rule out this behavior in the case of parallel junctions where one would be very close to a thermal regime at low temperatures.  In any case the unusual development of the spectra with temperature has prevented reasonable fits of the spectra taken at higher temperatures to the BTK formalism where the parameters $\Gamma_i$, $z_i$ and $\alpha$ should be kept constant once obtained from the fit to the data measured at the lowest temperature.

\begin{figure}[t]
\includegraphics[width=7 cm]{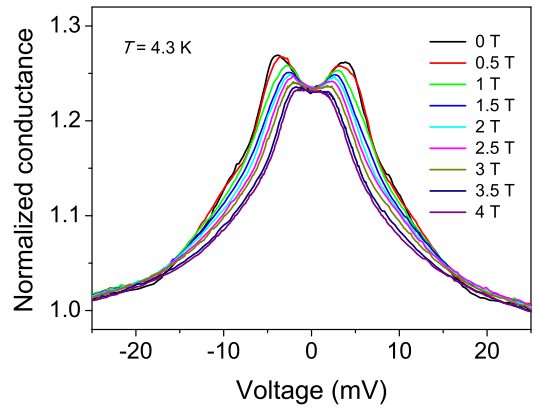}
\caption {PCAR spectra of the junction from Fig. 3a in magnetic fields.}
\label{fig:fig4}
\end{figure}

\begin{figure}[t]
\includegraphics[width=7 cm]{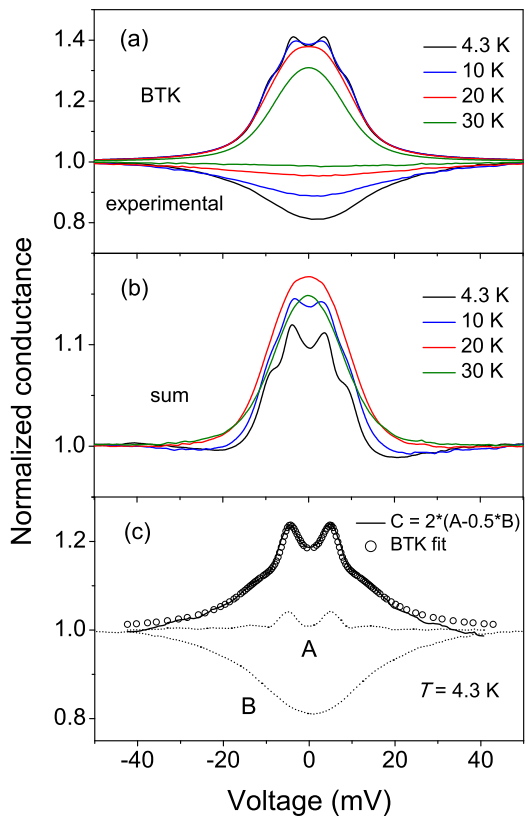}
\caption{Panel a: Upper curves - two-gap spectra generated by BTK model with parameters indicated in the text. Lower curves - set of measured tunneling-like spectra. Panel b: Sums of the two corresponding curves from upper panel. Panel c: Curve A - spectrum from Fig. 3b normalized to conductance background at higher voltages. B - spectrum from Fig. 5a. C - difference between A and B (see the text).}   
\label{fig:fig5}
\end{figure}

Due to presence of magnetism in the parent compound and/or magnetic impurities (Fe, Nd) possibly present in the sample also the Kondo scattering must be considered \cite{naidyuk}. The conductance minimum (or resistance maximum) around zero-bias voltage can be caused in principle also by the Kondo effect. Its appearance in the spectrum  can be tested by magnetic field. In increasing field the zero-bias-conductance peak should appear due to the Zeeman splitting  of degenerate spin states. Figure 4 presenting a behavior of the junction shown already in Fig. 3a, now in magnetic field up to 4 Tesla excludes the presence of this effect. As can be seen in the figure the gap-like peaks ($\sim \Delta_1$) and humps attributed to the large gap $\Delta_2$ are suppressed in a standard way as expected for the S/N junction with the superconductor S in the mixed state. It is because in this spatially inhomogeneous situation a full distribution of the order parameters stretching from zero at the vortex cores up to a maximal value determined by maximal internal field $B_i$ between the vortices is found. 

In the following we show that at least part of the unusual features found in the point-contact junctions here and also by other authors on the iron pnictides can be ascribed to the fact that parallel conductance  channels are realized by point-contact junctions on the polycrystalline samples.
In Fig. 5 an effect of the superposition of two parallel junctions, one with the two-gap spectrum and another showing the tunneling-like conductance, is presented. The two-gap spectrum (BTK )has been generated by the two-gap BTK formula with $ \Delta_1 = 5$ meV, $ \Delta_2 = 10$ meV,  $\Gamma_1 =$ 1 meV, $\Gamma_2 =$ 2 meV, $z = 0.4$ and $\alpha =$ 0.5. The tunneling-like spectra (experimental) are taken from experiment. The sum of these two spectra taken with equal weight is shown in Fig. 5b. Although not all details of the temperature evolution of the spectrum shown in Fig. 3a are reproduced one can notice that the conductance spectrum near the zero-bias voltage is increasing with the temperature  up to 20 K followed by its decrease at higher temperatures. It means that the unusual  increasing intensity of the PCAR spectrum with increasing temperature has not been necessarily connected with unconventional superconductivity in the system.

In  Fig. 5c the spectrum presented in Fig. 3b at 4.3 K is reproduced after normalization to its high-voltage background (curve A).  The normalized spectrum was impossible to fit by the BTK formalism. Let us consider again a parallel connection of two junctions, one with a two-gap spectrum and another with a tunneling-like character. Be the junction A represents the sum of these two. In Fig. 5c the conductance of the tunneling-like junction (curve B) was subtracted with a proper weight (0.5) from the conductance of the junction A. The resulting spectrum is indicated by the curve C. The spectrum C could be well fitted by the two-gap BTK formula. The circles represent this fit with $ \Delta_1 = 5.5$ meV, $ \Delta_2 = 11$ meV, $\Gamma_1 =$ 1.35 meV, $\Gamma_2 =$ 6 meV and $\alpha =$ 0.5. Using this simple model we have arrived to the two superconducting gaps which are  consistent with those found on the junctions A,B and C presented in Fig. 1. 

Recently the concept of the two-gap superconductivity in pnictides  has been supported by other works as the ARPES experiments of Ding {\it et al.} \cite{ding} on (Ba,K)Fe$_2$As$_2$. Also Gonnelli {\it et al.} \cite {gonnelli} have obtained pronounced two-gap PCAR spectra and pseudogap on the LaFeAsO$_{1-x}$F$_{x}$ polycrystals.

\section{CONCLUSIONS}

Systematic studies on the NdFeAsO$_{0.9}$F$_{0.1}$ superconductor show  an evidence of the two-gap superconductivity with the gap values $\Delta_1 = 5 \pm $1 meV and  $\Delta_2 = 11 \pm $2 meV indicating very weak coupling in the band with the small gap with $2 \Delta_1 /k_B T_c = 2.6 \pm 0.1$ and  strong coupling for the second band with $2 \Delta_2 /k_B T_c= 5.7 \pm 0.5 $. Also the indication for a reduced DOS in the normal state or pseudogap persisting well above the bulk transition temperature is found in the system.

\acknowledgments
This  work  has  been  supported  by  the Slovak Research and
Development Agency  under the contracts
Nos. APVV-0346-07, VVCE-0058-07 and LPP-0101-06, by the EC Framework Programme MTKD-CT-2005-030002.  Centre of Low Temperature Physics is
operated as  the Centre of Excellence  of the Slovak Academy
of Sciences.  The  liquid   nitrogen  for  the
experiment has  been sponsored by the  U.S. Steel Ko\v sice,
s.r.o.  
Work at the Ames Laboratory was
supported by the Department  of Energy - Basic Energy Sciences under  Contract No. DE-AC02-07CH11358.

\end{document}